\documentclass[
	pra,
	aps,
	reprint,
	amsmath,amssymb,
	a4paper,
	superscriptaddress,
	10pt,
    nofootinbib
]{revtex4-2}

\usepackage{graphicx}
\usepackage{physics}
\graphicspath{{fig/}}
\usepackage[dvipsnames,table,xcdraw]{xcolor}
\definecolor{rmpblue}{HTML}{2e3092}
\usepackage[
	colorlinks=true,
	citecolor=rmpblue,
	linkcolor=rmpblue,
	filecolor=magenta,
	urlcolor=rmpblue,
]{hyperref}
\usepackage[
	colorlinks=true
]{hyperref}
\usepackage{tikz}

\newcommand{\affilLOQM}{Laboratory of Optics of Quantum Materials, Department of Physics, Indian Institute of Technology Bombay, Mumbai 400076, India}

\newcommand{\affilQUNTIITB}{Centre for Excellence in Quantum Information, Computation, Science and Technology, Indian Institute of Technology Bombay, Powai, Mumbai 400076, India}

\newcommand{\affilHEUChina}{Qingdao Innovation and Development Center, Harbin Engineering University, Qingdao, Shandong, 266000, China}

\newcommand{\affilITMO}{School of Physics and Engineering, ITMO University, Saint-Petersburg, 197101, Russia}

\newcommand{\affilJNCASR}{School of Advanced Materials (SAMaT), Jawaharlal Nehru Centre for Advanced Scientific Research (JNCASR), Bengaluru, 560064, India}

\newcommand{\affilJNCASRChem}{Chemistry and Physics of Materials Unit (CPMU), Jawaharlal Nehru Centre for Advanced Scientific Research (JNCASR), Bengaluru, 560064, India}

\newcommand{\affilTIFR}{Department of Condensed Matter Physics, Tata Institute of Fundamental Research, Mumbai 400005, India}

\begin{document}
	
\title{Sensing with Broken Symmetry: Revisiting Bound States in the Continuum}
\author{Brijesh Kumar}
\affiliation{\affilLOQM}

\author{Elizaveta Tsiplakova}
\affiliation{\affilHEUChina}
\affiliation{\affilITMO}

\author{Samuel John}
\affiliation{\affilTIFR}

\author{Parul Sharma}
\affiliation{\affilLOQM}

\author{Nikolay Solodovchenko}
\affiliation{\affilHEUChina}
\affiliation{\affilITMO}

\author{Andrey Bogdanov}
\email{a.bogdanov@hrbu.edu.cn}
\affiliation{\affilHEUChina}
\affiliation{\affilITMO}

\author{Shriganesh S. Prabhu}
\email{prabhu@tifr.res.in}
\affiliation{\affilTIFR}

\author{Abhishek Kumar}
\email{abhishekkumar@jncasr.ac.in}
\affiliation{\affilJNCASR}
\affiliation{\affilJNCASRChem}

\author{Anshuman Kumar}
\email{anshuman.kumar@iitb.ac.in}
\affiliation{\affilLOQM}
\affiliation{\affilQUNTIITB}

\date{\today}
	
\begin{abstract}
Metasurface with bound states in the continuum (BICs) offer exceptional potential for optical sensing due to their inherently high quality (Q) factors. However, the detection of symmetry-protected BICs remains experimentally challenging due to their non-radiative nature. Introducing slight asymmetry makes these resonances observable, though it reduces the Q-factor. In real devices, intrinsic material losses further affect the resonance behavior and sensing performance. While it is often assumed that sensing is optimized at the critical coupling when radiative and non-radiative losses are balanced, the precise conditions for achieving the best limit of detection (LOD) and figure-of-merit (FOM) remain under active discussion. In this work, we experimentally and theoretically investigate BIC-based sensing in the terahertz (THz) range. We demonstrate that the LOD exhibits a non-monotonic dependence on asymmetry, reaching an unexpected optimum where radiative and non-radiative losses are not equal. Moreover, we show that this optimum differs between reflection and transmission sensing schemes. Our results provide practical guidelines for optimizing Q-factor, sensitivity, and signal amplitude together, and contribute to a deeper understanding of the fundamental limits of BIC-based sensing.
\end{abstract}   
   
\maketitle

\section{Introduction}
\textcolor{black}{Recent developments in nanophotonics have established metasurfaces as powerful platforms for advancing optical sensing technologies \cite{Shiue2017}. The phase, amplitude, and polarization of light can be precisely controlled with these metasurfaces, which are planar arrays of subwavelength structures \cite{Yu2014_meta,Kildishev2013_meta,Caldarola2015_meta}. These subwavelength scale structures provide new means of enhancing light-matter interactions, concentrating electromagnetic fields, and customizing optical resonances—all of which are essential for high-sensitivity sensing \cite{beruete2020terahertz:sens,zhang2020metasurfaces:sens,karawdeniya2022surface:sens,wang2023metasurface:sens}. Because of this, metasurfaces are being investigated extensively for uses where spectral selectivity, sensitivity, and compactness are crucial.}
\begin{figure*}
\centering
\includegraphics[width=\textwidth]{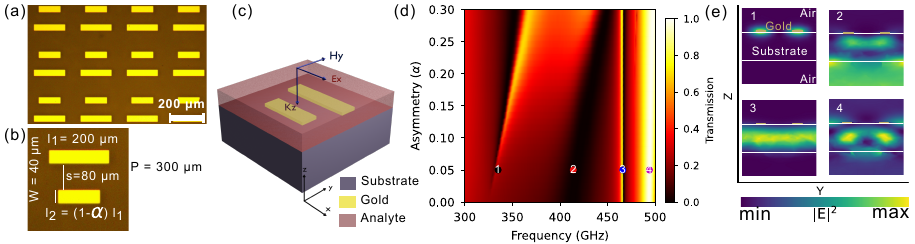}
\caption{$\mid$ {\bf  THz metasurfaces with q-BICs.} (a), Optical microscope image of the fabricated sample.  (b) A unit cell of metasurface. (c) Schematic of sensor.   (d) Simulated transmission of designed metasurface as an asymmetry function. (e) Intensity distribution for the modes marked by numbers in panel (d).}\label{fig1}
\end{figure*}
Recent progress in photonics has created new possibilities for improving sensing technologies through the use of metasurfaces\cite{khan2022optical:sens,la2019metasurfaces:sens,qin2022metasurface:sens,kazanskiy2022recent:sens}. 
One promising approach involves bound states in the continuum (BICs), which have attracted growing interest due to their ability to trap light with very low energy loss \cite{Hsu2016BIC,Hsu2013BIC,Romano2018BIC}. A well-known type of BIC is the symmetry-protected (SP) BIC, which appears in dielectric and plasmonic metasurfaces when it follows certain symmetry rules \cite{Jia2023sym,Zhen2014sym,Tsimokha2022sym,Koshelev2020sym}. By slightly changing the symmetry, SP-BICs can be turned into quasi-BICs. These modes still offer high and adjustable quality factors (Q factors) and can concentrate \textcolor{black}{electromagnetic} fields strongly in small regions, making them highly suitable for sensitive and precise sensing, including applications in biosensing~\cite{dong2024:QBICSens,peng2024terahertzQBICSens,luo2024qbics,tan2021qbics,qian2024qbics}.

\textcolor{black}{According to perturbation theory, sensor sensitivity is governed by the overlap integral of the electric field with the perturbed volume introduced by the analyte, emphasizing the importance of field localization in the sensing region}~\cite{teraoka2006:Perterbation,wang2021}. These theoretical developments primarily relate the intrinsic properties of the system. Specifically, the electric field \textcolor{black}{localisation} of the eigenmodes to sensor performance, assuming no external excitation. However, for practical sensor design, the coupling of the excitation field to the metasurface must also be considered. \textcolor{black}{When modes are excited via an external electromagnetic field, the resulting measurable response, such as transmission or reflection at resonance, becomes an important factor for sensing applications.} Recent studies have emphasized the critical role of amplitude in determining the actual performance of a sensor~\cite{Conteduca:acsphotonics2022:Amplitudeimportance,cong2015fano}. Therefore, to capture the full sensing response, we extend our definition of the figure of merit (FoM) to include not only the Q factor and field enhancement but also the resonance amplitude. This comprehensive approach provides a more accurate framework for evaluating and optimizing sensor performance.

In this work, we investigate the structural conditions that maximize the figure of merit (FoM) and sensitivity in metallic metasurfaces operating near quasi-bound states in the continuum (qBIC) for \textcolor{black}{thickness sensing in THz spectral range}. Our study emphasizes the practical importance of optimizing FoM, which incorporates both the sensitivity and resonance amplitude. To optimize the FoM of the qBIC sensor, we consider realistic  losses introduced by the substrate as well as the metal in our analysis. We use quasinormal mode (QNM) analysis to extract modal parameters from numerically simulated transmission spectra and validate our findings through experimental measurements on fabricated metallic metasurfaces consisting of asymmetric pair-rod scatterers. Metallic metasurfaces offer strong field confinement at the surface, which enhances interaction with the analyte and improves sensing performance compared to dielectric platforms. \textcolor{black}{Counterintuitively, our results reveal that maximum FoM is achieved near a critical coupling condition, where radiative and nonradiative losses are not balanced.} This work highlights the role of amplitude, asymmetry, and loss in shaping sensing performance and demonstrates that maximizing FoM provides a more practical design strategy for metallic metasurface sensors than conventional LoD-based optimization.

\section{Results and Discussion}\label{sec2}
\subsection{Existence of Symmetry-Protected Bound States in the Continuum (SP-BIC) in Metasurface Design}

We design a metasurface composed of a periodic array of two parallel gold bars on a dielectric substrate. The unit cell of this structure exhibits \( C_{2v} \) point group symmetry, which includes four symmetry operations: the identity (\( E \)), a 180$^0$ rotation about the \( z \)-axis (\( C_2 \)), and two mirror reflections across the \( x \) and \( y- \)axes, denoted as \( \sigma_x \) and \( \sigma_y \), respectively. The \( C_{2v} \) point group has four irreducible representations: \( A_1 \), \( A_2 \), \( B_1 \), and \( B_2 \), which describe how modes transform under symmetry operations. Among these, \( B_1 \) and \( B_2 \) correspond to \( x \)- and \( y \)-polarized fields, while \( A_1 \) and \( A_2 \) are symmetry mismatched with the far-field polarization.



In the far field, the electric field must be transverse, and is expressed as \( \mathbf{\Vec{E}} = (E_x, E_y) \), where \( E_x \) and \( E_y \) are the in-plane components. These components transform under the \( C_{2v} \) symmetry operations according to the \( B_1 \) and \( B_2 \) irreps, corresponding to \( x \)- and \( y \)-polarized light, respectively~\cite{tsimokha2022acoustic}. Consequently, only those eigenmodes that transform as \( B_1 \) or \( B_2 \) can couple to the corresponding polarization in the far field. Modes transforming under the \( A_1 \) or \( A_2 \) irreps are symmetry mismatched with the far-field polarization and thus remain decoupled. These uncoupled modes are known as \textbf{symmetry-protected bound states in the continuum (SP-BICs)}.

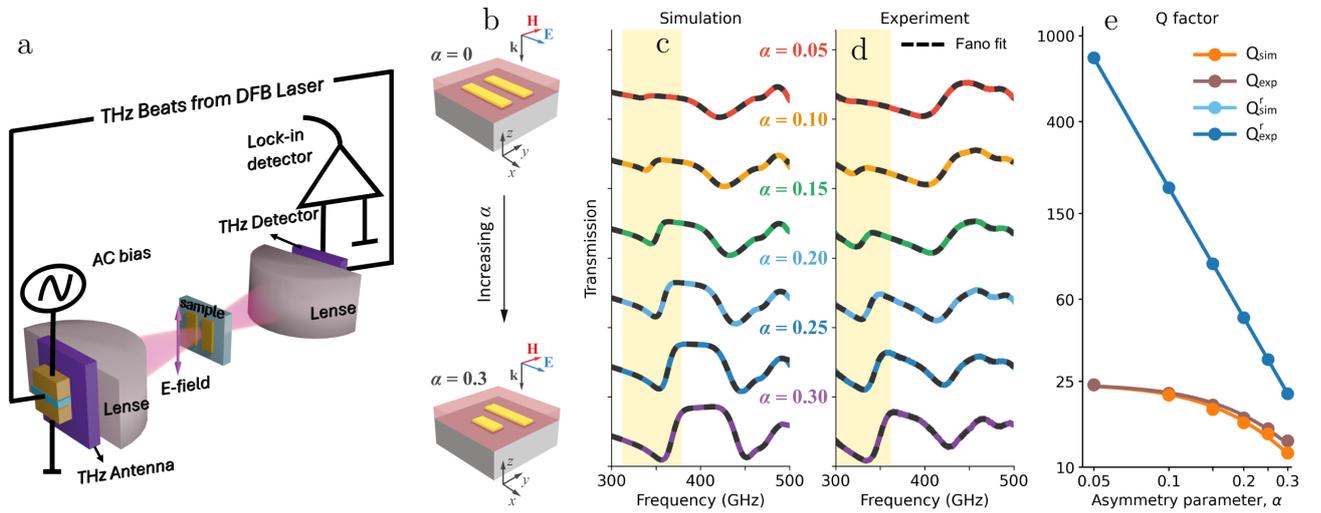
\begin{figure*}
\centering
\begin{tikzpicture}
\node[anchor=north west,inner sep=0] at (0,0)
{\includegraphics[width = 0.95\textwidth]{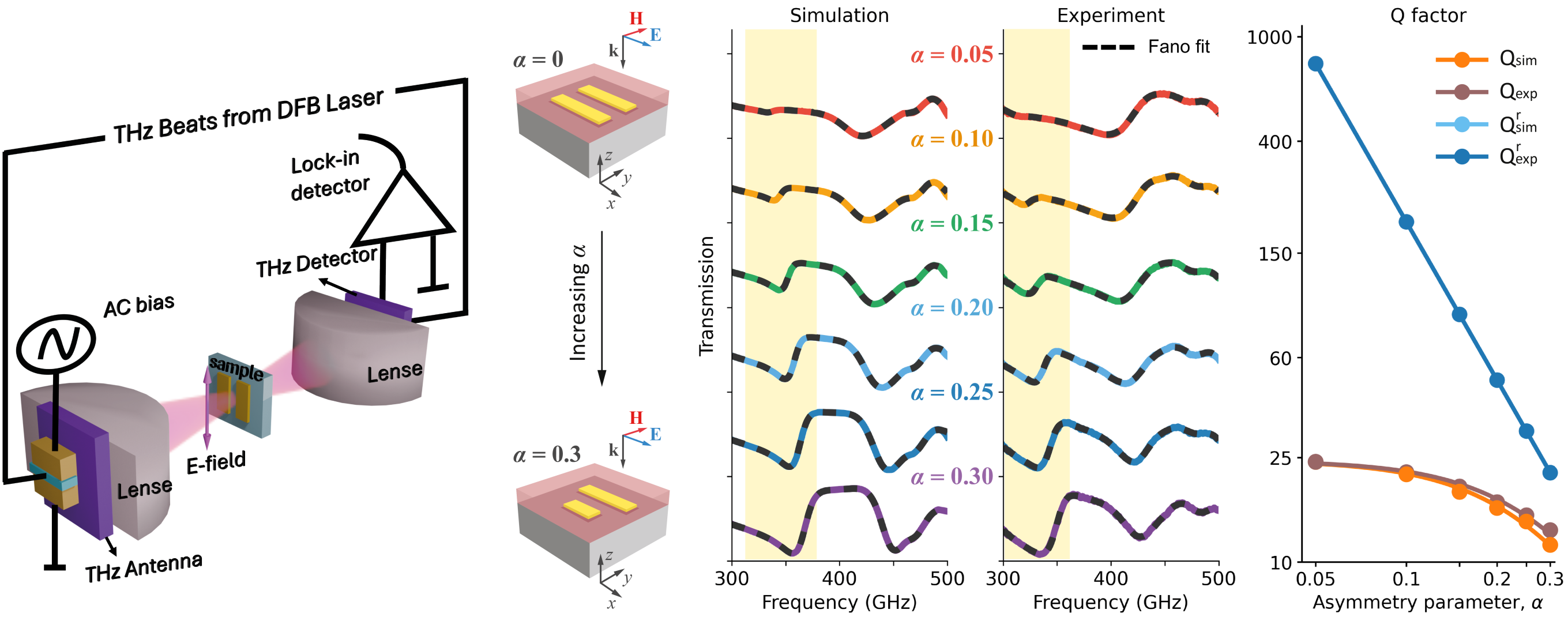}};
\small
\draw (0,-0.3) node [below right=1.25mm, circle, inner sep=0.1mm, fill = white]{\large{a}};
\draw (6.1,0.1) node [below right=1.25mm, circle, inner sep=0.1mm, fill = white]{\large{b}};
\draw (8.4,-0.3) node [below right=1.25mm, circle, inner sep=0.1mm]{\large{c}};
\draw (10.95,-0.3) node [below right=1.25mm, circle, inner sep=0.1mm]{\large{d}};
\draw (14.3,0) node [below right=1.25mm, circle, inner sep=0.1mm]{\large{e}};
\end{tikzpicture}
\caption{$\mid$ (a) Schematic of experimental setup. (b) Evolution of unit cell asymmetry in the fabricated metasurfaces. (c) Simulated and (d) experimentally measured transmission spectra (without analyte) corresponding to the structural variations in (b). The black dashed lines represent Fano-resonance fits. (e) Q-factor as a function of asymmetry.}
\label{fig:Qfactor analysis}
\end{figure*}

\textcolor{black}{Fig.~\ref{fig1}(a) shows the optical image of the fabricated quasi-bound state in the continuum (qBIC) metasurface, comprising two rectangular gold scatterers. The geometry of the unit cell, including all relevant dimensional parameters, is detailed in Figure 1b. The metasurface is fabricated on a quartz substrate, and structural asymmetry is introduced by varying the length of one of the antennas, quantified by the asymmetry parameter $\alpha$. In realistic scenarios, material dispersion and losses are inevitable. To account for these effects, the experimentally measured permittivity of the quartz substrate is incorporated in our analysis. Additionally, Ohmic losses in the gold antennas are modeled using a surface conductivity approach. The metasurface supports multiple modes, among which \textit{mode 1}, appearing at $320\ \mathrm{GHz}$, corresponds to a symmetry-protected qBIC. Upon breaking the in-plane symmetry, these modes begin coupling to the far field, transforming into quasi-BICs that retain high, yet finite, quality ($Q$) factors. Besides the qBIC, other modes are also marked in Fig.~\ref{fig1}(d), and their respective simulated electric field profiles are shown in Fig.~\ref{fig1}(e). Notably, the qBIC exhibits strong electric field confinement at the surface of metal scatterers.}



\subsection{Metasurface Fabrication and Optical Charectrisations}
The fabrication of a metasurface starts with substrate cleaning followed by optical lithography patterning. Gold was deposited by sputtering, and the liftoff process was performed to get the final structure. The transmission spectrum was then measured by continuous-wave (cw) terahertz spectroscopy, shown in Fig~\ref{fig:Qfactor analysis}(a). We prepared samples with varying geometrical asymmetry ($\alpha$) \textcolor{black}{defined as $\alpha = 1-{l_2}/{l_1}$} values ranging from 0 to 0.35, as depicted in Fig~\ref{fig:Qfactor analysis}(b). The simulated spectrum spectrum, shown in Fig.~\ref{fig:Qfactor analysis}(c), closely matches with the experimental transmission  in Fig.~\ref{fig:Qfactor analysis}(d), though the experimental resonance shows a slight redshift compared to the simulation. Additionally, we fitted the resonance near 0.325 THz with a Fano lineshape using quasi-normal mode theory to obtain the $Q$-factor of the peak  \cite{QNM_theory,Lalanne2018QNM}, whose variation with asymmetry is shown in Fig.~\ref{fig:Qfactor analysis}(e).


There are two dominating loss mechanisms i.e radiative loss and intrinsic loss (including absorption). The $Q_\text{tot}$ obtained from the spectrum depends on the radiative Q-factor ($Q_\text{rad}$) and intrinsic Q-factor ($Q_\text{int}$)  as: 
\begin{equation}
    \frac{1}{Q_\text{tot}} = \displaystyle\frac{1}{Q_\text{rad}} + \frac{1}{Q_\text{int}}
    \label{eq:Qt}
\end{equation}
where $Q_\text{rad}$ depends on the asymmetry parameter as \cite{Koshelev2018},
\begin{equation}
    Q_\text{rad} = Q_0\alpha^{-2}
\end{equation}
Where \(Q_0\) is a constant determined by the metasurface design, being independent on asymmetry \(\alpha\) leading to the relation \cite{Koshelev2019},
\begin{equation}
    \displaystyle Q_\text{tot} = \displaystyle\frac{Q_\text{int}}{x^2+1} \label{eq:Qt_alpha_cr}
\end{equation}
Here \(x = \alpha/\alpha_\text{cr} \) and $\alpha_\text{cr}$ is the critical value of asymmetry parameter defined as $\alpha_\text{cr} = \sqrt{{Q_{0}}/{Q_\text{int}}}$, which represents the value of the asymmetry parameter at which radiative loss equals intrinsic loss. In practical applications, losses cannot be eliminated but can only be minimized. Therefore, $\alpha_\text{cr}$ is a crucial parameter when utilizing the qBIC device in real-world scenario.
\begin{figure*}
\centering
\includegraphics[width=\textwidth]{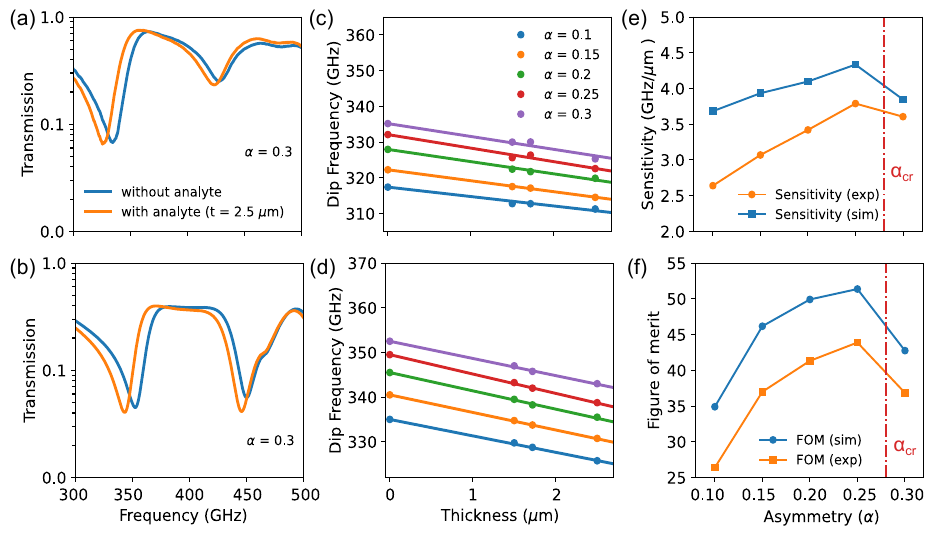}
\caption{$\mid$ {\bf BIC and q-BIC analysis:-} {\bf a}, Transmission of asymmetry $\alpha =0.3$ with (orange) and without (blue) analyte from FEM simulation  {\bf b},  Transmission of asymmetry $\alpha =0.3$ with (orange) and without (blue) analyte from experimental measurement, {\bf c, d}, calculation of sensitivity by fitting the thickness vs resonance shift from simulation and experimental, respectivly, {\bf e}, sensitivity with the function of asymmetry {\bf f}, Figure of merit with the function of asymmetry }\label{fig3}
\end{figure*}

\subsection{Asymmetry-Driven Sensitivity and Figure of Merit Analysis} 
To characterise a sensor, the key characteristics typically include sensitivity, detection limit, and the figure of merit (FoM). These parameters provide a comprehensive understanding of the sensor performance and its ability to detect and respond to changes in the environment. In this section, we examine the sensitivity and FoM of our qBIC sensor as a function of the asymmetry parameter. First, we analyze the sensor response to changes in thickness by placing a analyte of refractive index ($n_a$ = 1.67)  \cite{Pan2019} on the metasurface and varying its thickness ranging from 1.5 $\mu$m to 2.5 $\mu$m. The sensitivity $\mathcal{S}$ is defined as 
\begin{equation}
    \mathcal{S} = \left(\frac{\partial f_\text{qBIC}}{\partial t}\right)_{t=t_0}
    \label{eq:sens}
\end{equation}
where $t$ denotes the thickness of the analyte layer. 

In far-field measurements, the minimum observable transmission peak limits the sensitivity at lower asymmetry values. For experimental sensitivity measurements, a multi-layer thin film of photoresist (S1813) was spin-coated on the metasurface, with thickness varying from ${1.5}$ to $2.5 \ \mu$m measured by \textcolor{black}{DektakXT profilometer (Bruker)}. The resonance frequency of each spectrum was extracted by choosing the frequency at minima of the spectrum, and the shift was calculated by subtracting the bare metasurface spectrum (\textcolor{black}{i.e., without analyte}). Figure~\ref{fig3}(a) and~\ref{fig3}(b) show the FEM-simulated and experimentally measured transmission spectra, respectively, for structures with and without the analyte layer. In both cases, we set \(\alpha =0.3\) to ensure a consistent comparison. The presence of the analyte induces a shift in the resonance dip, which is clearly visible in both simulations and measurements. To quantify this shift, Figs.~\ref{fig3}(c) and ~\ref{fig3}(d) present the dependence of the resonance frequency dip on analyte thickness for various values of \(\alpha\). The x-axis represents the analyte thickness, while the y-axis shows the corresponding resonance frequency. Each dataset is fitted with a straight line, and the slope of these lines indicates the sensor's sensitivity, defined as the frequency shift per unit thickness of the analyte as defined in Eq.(\ref{eq:sens}). 
Figure~\ref{fig3}(e) presents the variation of sensitivity as functions of the asymmetry parameter $\alpha$. Which includes two curves—one based on FEM simulations and the other from experimental measurements. The sensitivity values are extracted from the slopes of the linear fits shown in Figs.~\ref{fig3}(c) and ~\ref{fig3}(d). These results illustrate how sensitivity improves with increasing asymmetry, highlighting the tunability of the sensing performance through quasi-bound states in the continuum (qBIC). Notably, the sensitivity does not increase indefinitely; instead, it reaches a maximum at an optimal asymmetry value.

For qBIC sensors, it is typically the case that the strength of the resonance intensity is inversely correlated with the quality factor \cite{cong2015fano}. \textcolor{black}{Where a large Q factor is necessary in order to resolve the  fractional frequency shift \(\Delta\omega/\omega\) predicted by the perturbation theory \cite{Zhang2014}}
\begin{equation}
    \frac{\Delta\omega}{\omega_0} \propto \frac{\int_{V_p}\Delta\varepsilon\left|\mathbf{E}\right|^2 \text{d}^3\textbf{r}}{\int_{V}\mathbf{E}^{*}\cdot\varepsilon\  \mathbf{E}\text{d}^3\textbf{r}}
    \label{Eq:sens-by-perturbation}
\end{equation}
Here $V_p$ is the volume of perturbation and $V$ is volume of metasurface unit; $\Delta \omega$ is the change in frequency due to perturbation of permittivity $\Delta\varepsilon$ and $\omega_0$ denotes the resonance frequency of the unperturbed metasurface. From the equation, it is clear that the sensitivity of a metasurface-based sensor increases with electric field confined near the surface \cite{wang2021}
For practical sensors, however, the peak strength is also essential as it indicates the prominence of the transmission dip above the noise level, which is crucial for reliable measurements. \textcolor{black}{The impact of losses on the resonance amplitude is significant, as it limits the observable signal in far-field measurements. Using coupled mode theory (CMT) (see Supplementary Information), the transmission resonance amplitude ($\mathrm{h}$) in the presence of losses can be described as follows:}
\begin{equation}
    {h}(\lambda_0) = \left( \displaystyle\frac{Q_{\mathrm{tot}}}{Q_{\mathrm{int}}} \right)^2 
\end{equation}
\textcolor{black}{This ratio captures the relative contribution of intrinsic losses and directly impacts the strength of the resonance in transmission. As a result, The expression for the minimum detectable frequency deviation $ \Delta\omega_{\min}$  in transmission is given as (see Supporting Information for the full derivation)}
\begin{equation}
    \Delta\omega_{\min} \approx \displaystyle\frac{\omega_0}{Q_{\mathrm{int}}\frac{x\sqrt{2+x^2}}{(1+x^2)^2}}\sqrt{3\sigma}
    \label{eq:lam_min}
\end{equation}
\textcolor{black}{where $3\sigma$ defined as  minimum detectable amplitude variation and $x=\alpha/\alpha_{\text{cr}}$. Equation~(\ref{eq:lam_min}) can now be easily correlated with the minimum detectable change in the measurand through the sensitivity $\mathcal{S}$. Thus, the  limit of detection ($\mathrm{LoD}$) is given by
} \cite{Conteduca2022}
\begin{equation}
    \mathrm{LoD} = \frac{(1+x^2)^2}{\mathcal{S}\cdot Q_{\text{int}}x\sqrt{2+x^2}}\omega_0 \sqrt{3\sigma}
    \label{eq:LoD}
\end{equation}

Thus we define the figure of merit $\mathrm{FoM}$ of a qBIC-based sensor in far-field measurement as
\begin{equation}
    FoM = \mathcal{S}\cdot Q_{\text{int}}\frac{x\sqrt{2+x^2}}{(1+x^2)^2}
    \label{FoM-eq}
\end{equation}
Maximizing this FoM, which is equivalent to minimizing the LoD. In our approach, simulated and experimentally measured FoM is plotted in Fig.~\ref{fig3}(f), which shows a consistent trend as a function of asymmetry. \textcolor{black}{Importantly, we find that this figure of merit is different for reflection based and transmission based qBIC sensors, as shown in the supplementary information.} 

If we consider the sensitivity \( \mathcal{S} \) to be independent of the asymmetry parameter \( \alpha \), the figure of merit (FoM) is maximized at \( \alpha_{\text{cr}}\sqrt{\sqrt{2} - 1} \approx 0.64\alpha_{\text{cr}} \). However, since in our case the sensitivity depends on \( \alpha \), the FoM is instead maximized within a range between  \( 0.66\alpha_{\text{cr}}\) and \(  \alpha_{\text{cr}} \). This relationship, to the best of our knowledge, has not been explored in the literature. This behavior is consistent with our findings in Fig.~\ref{fig:Qfactor analysis}, where the maximum FoM is observed around the asymmetry parameter \(  0.66\alpha_{\text{cr}} <\alpha<\alpha_{\text{cr}}\) in simulation as well as experimental data. Correspondingly, optimization of our experimental FoM corresponds to the minimum limit of detection of the qBIC sensor. Our experimental results show that it is important to consider the influence of nonradiative losses in the substrate as well as the antennae while optimizing the design of qBIC sensors, and that there is a non-trivial dependence of the limit of detection on the asymmetry parameter. 

 \section{Conclusion}
{In this work, we have investigated the role of structural asymmetry in optimizing the sensing performance of metallic metasurfaces supporting qBICs. By systematically tuning the asymmetry parameter, we demonstrated both experimentally and through simulations that sensitivity and FoM exhibit non-monotonic behavior, with the LOD reaching an optimal value near an optimal asymmetry \textcolor{black}{where counterintuitively, radiative and nonradiative losses are not balanced.}

{Using quasinormal mode theory and perturbation analysis, we incorporated the effects of electric field confinement, resonance linewidth, and amplitude to develop a comprehensive figure of merit that directly correlates with the sensor's limit of detection. Our results show that the interplay between radiative quality factor and intrinsic losses significantly impacts the observable resonance amplitude and thus the practical sensitivity.}

{We validated our theoretical model using fabricated metasurfaces and refractive index sensing experiments with a thin analyte layer. The excellent agreement between experimental and simulated trends confirms that while the optimal performance depends on the critical coupling, the optimum asymmetry value occurs in a window below $\alpha_{\text{cr}}$. We have presented a recipe for such an optimization and experimental validated the same via terahertz transmission measurements. These findings underscore the importance of considering all loss mechanisms as well as resonance amplitude in the design of qBIC-based sensors.}
{Our study establishes a robust framework for optimizing metasurface sensors by jointly maximizing Q-factor, sensitivity, and amplitude, and provides key insights for the development of highly sensitive, low-LoD optical sensors.}



\begin{acknowledgements}
A.K. acknowledges funding support from the Department of Science and Technology via grant number CRG/2022/001170. B.K. acknowledges support from Prime Minister's research fellowship (PMRF), Government of India. We acknowledge IITBNF and SAIF/CRNTS at IIT Bombay for providing fabrication and characterization facilities for this work. The authors declare no competing financial interests.
\end{acknowledgements}

\bibliography{references}

\end{document}